# RANGE MARGIN REDUCTION IN CARBON ION THERAPY: POTENTIAL BENEFITS OF USING RADIOACTIVE ION BEAMS

Olga Sokol[1], Laura Cella[2], Daria Boscolo[1], Felix Horst[1,3], Caterina Oliviero[4], Roberto Pacelli[4], Giuseppe Palma[5], Micol De Simoni[6,7], Manuel Conson[4], Mara Caroprese[4], Ulrich Weber[1], Christian Graeff[1,8], Katia Parodi[7], Marco Durante[1,8*]

[1]GSI Helmholtzzentrum für Schwerionenforschung, Darmstadt, Germany

[2]Institute of Biostructures and Bioimaging, National Research Council (CNR), Napoli, Italy

[3]OncoRay - National Center for Radiation Research in Oncology, University Hospital Carl Gustav Carus, Technische Universität Dresden, Helmholtz-Zentrum Dresden-Rossendorf, Dresden, Germany

[4]Azienda Ospedaliera Universitaria - Federico II, Napoli, Italy

[5]Institute of Nanotechnology, National Research Council (CNR), Lecce, Italy

[6]INFN Section of Rome 1, Rome, Italy

[7]Department of Medical Physics, Ludwig-Maximilians- Universität München (LMU) Munich, Germany

[8]Technische Universität Darmstadt, Darmstadt, Germany

[*]Corresponding author, m.durante@gsi.de

## ABSTRACT

Radiotherapy with heavy ions, in particular, $^{12}$C beams, is one of the most advanced forms of cancer treatment. Sharp dose gradients and high biological effectiveness in the target region make them an ideal tool to treat deep-seated and radioresistant tumors, however, at the same time, sensitive to small errors in the range prediction. Safety margins are added to the tumor volume to mitigate these uncertainties and ensure its uniform coverage, but during the irradiation they lead to unavoidable damage to the surrounding healthy tissue. To fully exploit the benefits of a sharp Bragg peak, a large effort is put into establishing precise range verification methods for the so-called image-guided radiotherapy. Despite positron emission tomography being widely in use for this purpose in $^{12}$C ion therapy, the low count rates, biological washout, and broad shape of the activity distribution still limit its precision to a few millimeters. Instead, radioactive beams used directly for treatment would yield an improved signal and a closer match with the dose fall-off, potentially enabling precise *in vivo* beam range monitoring. We have performed a treatment planning study to estimate the possible impact of the reduced range uncertainties, enabled by radioactive $^{11}$C beams treatments, on sparing critical organs in the tumor proximity. We demonstrate that (i) annihilation maps for $^{11}$C ions can in principle reflect even millimeter shifts in dose distributions in the patient, (ii) outcomes of treatment planning with $^{11}$C beams are significantly improved in terms of meeting the constraints for the organs at risk compared to $^{12}$C plans, and (iii) less severe toxicities for serial and parallel critical organs can be expected following $^{11}$C treatment with reduced range uncertainties, compared to $^{12}$C treatments.

## I. INTRODUCTION

Charged particle therapy with protons or heavy ions (e.g., $^{12}$C) is nowadays considered to be the most advanced form of radiotherapy, offering new opportunities for non-invasive tumor treatments [1–4]. In contrast to the conventional radiation therapy with photons, it offers a more favorable dose distribution (the so-called Bragg peak) and is, therefore, better suited for the treatment of deep-seated tumors. Most of the deposited dose would be efficiently concentrated in the target, while the healthy tissue located in the plateau region would receive less damage. In addition, heavy ions offer increased efficacy compared to proton beams in terms of the high relative biological effectiveness (RBE) at the end of their range, meaning that for at a given absorbed dose, the biological effects would be 2-3 times higher [5]. However, in reality, the sharp dose fall-off is a double-edged sword as it makes the success of treatment extremely sensitive to the precise localization of the Bragg peak in the patient: under- or overestimations of the particle range would lead to the underdosage in the tumor and high-dose hotspot in the nearby healthy tissue. The uncertainties affecting the Bragg peak localization typically include the setup (patient positioning) and range (see below) errors, as well as uncertainties in beam delivery and dose calculation.

The range uncertainties in treatment planning come from unexpected variations of tissue density in the beam path (i.e., anatomical changes) or from the inaccuracies in the conversion of the CT Hounsfield units (HU) into the water-equivalent path length (WEPL) of the ions. Overall, they might alter the range of particles on the order of several millimeters. Accounting for the range uncertainty by either expanding the tumor planned volume (PTV) during the non-robust treatment planning procedure or by considering it in robust planning on a clinical target volume (CTV) would in both cases lead to the irradiation of healthy tissue and possibly critical structures in the target proximity, potentially inducing toxicities. For serial organs at risk (OAR), a dose higher than the tolerance threshold to any of its subunits is sufficient to increase the probability of a severe side effect significantly. Thus, one would expect that reduced tumor margins, associated with range uncertainty, significantly improve the treatment related toxicity. In fact, the decrease of head and neck toxicities was already predicted following the *in-silico* range margin reduction study in proton patients [6]. With regard to parallel OAR, where the damage would be correlated with a certain volume or the fraction of the organ receiving a threshold dose, the impact of the margin reduction is also expected to play a positive role however the size of the effect would probably strongly depend on the volumes of the OAR and the tumor.

To fully exploit the benefits of a sharp Bragg peak, a large effort is put nowadays into investigating different range verification methods. Among them, positron emission tomography (PET) remains one of the most widely used approaches to monitor dose delivery in clinical practice by exploiting $\beta^+$-emitting isotopes [7]. In $^{12}$C ion therapy, the contributors to the PET signal are the radioactive fragments of the primary ions, mainly $^{11}$C and $^{10}$C. The activity peak is then observed upstream of the Bragg peak because these fragments have shorter ranges than the primary projectiles at the same velocity [8]. Unfortunately, the usefulness of PET in $^{12}$C therapy remains marginal and insufficient for a complete range uncertainty elimination for several reasons. First, the half-life of the most abundantly produced radionuclides is too long for instantaneous feedback. Instead, the short-lived radionuclides are produced at a low rate and exhibit a long positron range before annihilation [9]. Second, the spatial shift in the $\beta^+$-activity and dose peaks as well as the biological washout requires Monte Carlo (MC) simulations [10] to analyze the data. Uncertainties in those calculations together with the low counting rates of β+-emitting fragments currently limit the accuracy of the PET-based range verification method to about 2-5 mm [11,12].

One could overcome the above-mentioned difficulties by using radioactive ion beams (RIBs), such as $^{11}$C or $^{10}$C for simultaneous treatment and online beam verification [13]. This way, compared to the irradiation with stable ions, all the primaries will potentially contribute to the PET signal, significantly increasing the signal-to-noise ratio [14]. Shorter half-lives of the RIBs (20.33 min and 19.3 s for $^{11}$C and $^{10}$C, respectively), would reduce the time of PET signal acquisition and make the impact of the biological washout less significant [15]. Finally, the activity peaks would match the mean ranges of the primaries and will be correlated with the R80 of the Bragg peaks (SOBP), i.e., the position in the distal fall-off where the dose drops to 80% of the SOBP value [16].

Until now, the pre-clinical and clinical studies of RIBs were held back by the difficulties in producing RIBs at sufficiently high intensities. However, the interest in these studies has been recently revived following the new developments at modern accelerator facilities, such as the intensity upgrade of the SIS-18 accelerator at GSI (Darmstadt, Germany) in frames of the FAIR facility construction [17]. The work on RIBs characterization and clinical application has been resumed in frames of the BARB (Biomedical Applications of Radioactive ion Beams) project [18] funded by the European Research Council (ERC) in 2020. There have been already two successful experimental campaigns in 2021 aimed at the RIB characterization and first imaging tests with PET detectors, however, these are not part of this contribution and will be presented in separate contributions [19].

In this study, instead, we attempt to demonstrate the rationale for using RIBs in clinical practice. First, we estimate the potential accuracy of dose visualization and its shift detection with RIB activity data. Following that, we perform an *in-silico* treatment planning study to qualitatively estimate the potential benefit from $^{11}$C treatment planning with reduced range margins, granted by the precise PET imaging of the beam. By retrospectively robustly re-optimizing patient plans for the head-and-neck and liver tumors with $^{12}$C and $^{11}$C beams, we estimate the impact of range margin reduction on serial (optic nerves) and parallel (liver) OAR on the level of treatment planning (i.e., meeting the dose constraints) and, furthermore, in terms of the predicted normal tissue complication probabilities (NTCP) [20].

## II. MATERIALS AND METHODS

### A. Treatment planning

1. Patient data

For this study, we have retrospectively considered consecutive patients with adenoid cystic carcinoma (ACC) and liver tumors to evaluate the impact of margin reduction on the damage to the serial (optic nerves) and parallel (liver) organs, respectively.

ACC patients (N=15) analysed here had tumors located mainly in the frontal areas of the skull and were originally treated in 2002-2008 at GSI in frames of the pilot project, implying a $^{12}$C ion boost following a course of intensity-modulated radiotherapy in Heidelberg [21,22].

The liver patients (N=10) studied here have originally undergone photon stereotactic body radiation therapy (SBRT) with abdominal compression at the Department of Radiotherapy of the University "Federico II" of Naples.

2. Planning considerations

GSI in-house treatment planning system TRiP98 for active beam scanning was used to perform biologically weighted robust treatment plan optimization [23,24] on a CTV using $^{12}$C and $^{11}$C beams. The treatment planning uncertainties included ±3.5% ($^{12}$C) or 0% ($^{11}$C) range uncertainty and ±3 mm shifts in all cardinal directions to consider positioning uncertainties. The robust optimization algorithm implicitly generates a safety margin to the CTV by considering several uncertainty scenarios during optimization. For each plan, a total of 21 uncertainty scenarios were analyzed.

The magnitudes of uncertainties for $^{12}$C were preserved from a previous work on robust optimization [24] (where they were adapted from the proton study [25]) for both sets of patient plans, even though for the liver with more homogeneous structure the range uncertainties might be slightly too conservative [26]. Target motion was not considered. The planning objective was to achieve at least 95% of the prescribed dose in at least 95% of the CTV volume in at least 95% of treatment scenarios, obeying the OAR constraints listed below. If a compromise could not be achieved, target coverage was favoured over OAR sparing.

*ACC*. We have preserved the original two-field configuration as well as the OAR contours from the original treatment plans from the pilot project, adjusting them to a 20 × 3 Gy(RBE) fractionation scheme. The planning constraint for the optic nerves was estimated following the European Particle Therapy Network (EPTN) recommendations [27] to $D_{0.03cm^3}$ < 2.46 Gy(RBE) per fraction. The $\alpha/\beta$ ratio of both tumor and normal tissues was assumed to be 2 Gy [27,28].

*Liver tumors*. Depending on the tumor location relative to the OARs, a single- or double-field configuration was chosen for each tumor. Where possible, a fractionation scheme of 4 × 10 Gy was used, otherwise, it was changed to 6 × 7 Gy to reduce the dose to the serial OAR in target proximity. Planning constraints were adapted to the fractionation schemes following the recommendations of the American Association of Physicists in Medicine group (AAPM) [29]. The $\alpha/\beta$ ratio of the tumor was assumed to be 10 Gy [30,31], and 2 Gy for the healthy liver [32,33].

The details for each treatment plan are given in Table I.

| Tumor | Patient ID | Target | $V_{CTV}$, cm$^3$ | Fractions | Fraction dose, Gy(RBE) | No. fields | OARs considered |
|---|---|---|---|---|---|---|---|
| ACC | 1 |  | 340.4 | 20 | 3 | 2 | Optic system (eyes, nerves chiasm), brainstem |
|  | 2 |  | 336.7 |  |  |  |  |
|  | 3 |  | 199.9 |  |  |  |  |
|  | 4 |  | 206.8 |  |  |  |  |
|  | 5 |  | 245.6 |  |  |  |  |
|  | 6 |  | 53.4 |  |  |  |  |
|  | 7 |  | 350.2 |  |  |  |  |
|  | 8 | a | 747.9 |  |  |  |  |
|  | 9 |  | 316.9 |  |  |  |  |
|  | 10 |  | 351.0 |  |  |  |  |
|  | 11 |  | 201.0 |  |  |  |  |
|  | 12 |  | 343.3 |  |  |  |  |
|  | 13 |  | 462.3 |  |  |  |  |
|  | 14 |  | 294.9 |  |  |  |  |
|  | 15 |  | 215.5 |  |  |  |  |
| Liver | 1 | a | 30.0 | 4 | 10 | 1 | Right kidney and lung |
|  | 2 | a | 16.1 | 4 | 10 | 1 |  |
|  | 3 | a | 17.4 | 4 | 10 | 1 |  |
|  | 4 | a | 12.2 | 6 | 7 | 2 | Small bowel, Colon |
|  | 5 | a | 70.6 | 4 | 10 | 1 |  |

| | | | | | | |
|---|---|---|---|---|---|---|
| 6 | a | 198.2 | 6 | 7 | 2 | Right lung, duodenum |
| | b | 24.6 | 6 | 7 | 1 | |
| 7 | a | 40.2 | 4 | 10 | 1 | |
| 8 | a | 162.2 | 6 | 7 | 1 | Large bowel, heart |
| 9 | a | 60.0 | 4 | 10 | 1 | Right kidney |
| | b | 38.0 | 4 | 10 | 1 | |
| | c | 32.8 | 4 | 10 | 1 | |
| 10 | a | 7.8 | 4 | 10 | 1 | |

*Table I. Adenoid cystic carcinoma (ACC) and liver patient plans characteristics summarizing the numbers and sizes of clinical tumor volumes, fractionation schemes (number and dose per fraction), numbers of fields, and organs at risk considered during the plan optimization.*

3. Base data

Apart from the patient-specific input (CT scans and RT structure contours, prescribed doses, and field configuration), TRiP98 requires a valid physical beam model describing the interaction of an ion with water, the reference medium in radiotherapy. The so-called base data includes laterally integrated depth-dose distributions in water and energy-dependent fragment spectra, necessary for the calculation of biological effects, at representative depths in water [34].

For this study, a new set of base data for $^{11}$C, which were not previously used in TRiP98, was generated with the FLUKA Monte-Carlo code (version 2020.1.10) [35], which was also used for calculating annihilation maps, see next section. To keep consistency between the two programs (FLUKA and TRiP98), energy loss tables for the primaries and all the fragments were extracted from FLUKA and used in TRiP98 together with the new set of base data. Additionally, still for consistency, a new set of base data for the $^{12}$C was also specifically generated for this work.

For both ion beams the FLUKA simulations were performed using the recommended default setting for ion beam therapy applications (cf. defaults 'HADROTHErapy'). For both isotopes, the beam spot was assumed to be Gaussian with a full width at half maximum (FWHM) of 8 mm in the isocenter. The Bragg peak broadening due to the use of a ripple filter was considered by the introduction of a momentum spread (dp/p) of 1% for all energies. Energy dependent fragment spectra and laterally integrated depth dose distributions were calculated in the energy range between 80 and 440 MeV/u, for beams impinging directly from vacuum to water, as in TRiP98 the energy degradation in the nozzle is considered through an adjustable offset value. A more in-depth description of the Monte Carlo calculation of base data for TPS can be found in [36].

B. Monte-Carlo simulation of annihilation maps for range shifts detection

To prove whether the information from the $^{11}$C β$^+$-decays can be at least theoretically used to precisely detect shifts between the planned and actual dose distributions, we have performed an additional TRiP98-FLUKA study in a patient geometry.

As a first step, we produced a treatment plan for a patient using the look-up tables for converting the HU into the ion's WEPL (Hounsfield look-up table, or HLUT) extracted from FLUKA. For simplicity, the plan was optimized for a uniform RBE-weighted dose of 3 Gy delivered with a single field. Then, the optimization procedure was repeated using the same HLUT scaled either by 1% or by 3.5% to emulate the range uncertainties associated with CT interpretation. As a result, three different raster scanner files containing information on energies and numbers of particles were produced. If loaded back into the system for the dose calculation with the default HLUT, only the first file would yield the dose distribution matching the target contours, while the other two would cause partial underdose and hotspots in the nearby healthy tissue.

In the second step, we loaded the patient CT and the raster scanner files in FLUKA to calculate the physical (absorbed) dose and annihilation distributions (as a best possible scenario the integral annihilation signal was scored). In this case the FLUKA built-in material description and HU-WEPL conversion were used for all three scenarios.

A specifically developed '*source.f*' user routine, similar to the one described in [37] was used to read the TRiP98 raster scanner files, specifying the sequence of the beam delivery, in FLUKA. This routine allows the possibility of full Monte-Carlo simulation of TPS-optimized plans while using the same beam settings adopted for the base data calculations. In addition to the calculation of the annihilation maps in a realistic treatment planning scenario, this routine was used for consistency check between the MC calculations and the expected dose deposition distributions obtained with TRiP98.

As a result, we have produced three sets of dose and annihilation maps that were visualized and analyzed.

### C. NTCP evaluation

All dose maps produced following ACC and liver patient plans optimization were first converted from Voxelplan (native TRiP98 format) into DICOM RT format using 3D Slicer [38] and then into Matlab-readable format (MathWorks, Natick, MA, USA) for further analysis. Next, each dose map has been voxelwise converted into 2-Gy equivalent dose (EQD2) by an in-house developed function for Matlab [39,40]. For the ocular structures in the ACC plans, the linear-quadratic model [41] was used with $\alpha/\beta$ = 2 Gy [28]. Instead, for the liver plans where high doses per fraction were planned (> 6 Gy per fraction), the EQD2 conversion was performed under linear-quadratic-linear condition [42], using the parameters $\alpha/\beta$ = 2.5 Gy, $\gamma/\alpha$ = 5 Gy, $d_t$ = 5 Gy as reported in [43,44].

NTCP values were estimated using the Lyman-Kutcher-Burman (LKB) model [45,46] with organ- and toxicity-specific parameters from the models reported in literature. For the optic nerves, only one model for the optic neuropathy endpoint was applied. Instead, for the healthy liver (liver minus GTV), several models related to endpoints of different severities exist and were included in the analysis. The models and their key parameters are summarized in Table II.

| Organ | Model | Endpoint | n (or a*) | m | TD$_{50}$, Gy |
|---|---|---|---|---|---|
| Optic nerve | Mayo [47] | Optic neuropathy | 0.25 | 0.14 | 72 |
| Healthy liver | Burman [48] | Liver failure | 0.32 | 0.15 | 40 |
| | Dawson [49] | Radiation-induced liver injury (grade ≥ 3) | 0.97 | 0.12 | 39.8 |
| | El Naqa I [43] | Changes in albumin-bilirubin (ALBI) and Child-Pugh (C-P) score | 1 | 1.47 | 24.3 |
| | El Naqa II [43] | Liver enzymatic changes (grade > 3) | 1 | 0.55 | 52.6 |

| Pursley I [44] | Change in C-P score (grade ≥ 2) | 0.06* | 0.8 | 19 |
| Pursley II [44] | Change in ALBI (grade ≥ 1) | 0.5* | 1.5 | 32 |

*Table II. Summary of the LKB NTCP model parameters used in the study. $TD_{50}$ is the value of the uniform dose given to the entire organ surface corresponding to the 50% probability to induce toxicity; m is inversely proportional to the slope of the dose-response curve; and n (or a) accounts for the volume effect.*

As pointed out by [44,50,51], the mean total (multiplied by the number of fractions) dose $D_{mean}$, as well as the V5 and V30 representing the fractions of the healthy liver receiving the total dose of more than 5 or 30 Gy, respectively, can be used as predictive parameters for estimating the hepatic toxicity.

### D. Statistical analysis

Wilcoxon signed rank test was used to study the differences between the $^{12}$C and $^{11}$C treatment planning and NTCP modelling outcomes and was performed with GraphPad Prism Software V9.4.0.

## III. RESULTS

### A. RIB annihilation maps to detect range shifts

Figure 1 shows the results of a single-field dose optimization on a patient CT. The RBE-weighted dose distribution, calculated with TRiP98, is shown in Figure 1a, while Figures 1b and 1c show the resulting absorbed dose and time-integrated annihilation maps calculated with FLUKA. In Figure 1d we include the beam-eye-view profiles of the absorbed dose and annihilation distributions yielded by the three raster files, produced using the default and scaled HLUT. Full distributions, normalized to the respective peak values, as well as the zoom-in of the target region are shown. The positions of the peaks of the annihilation distributions correlate with the positions of the fall-offs of the respective dose distributions, and the distances between them reflect the shifts in the dose distributions caused by either 3.5% or 1% differences in interpreting CTs in terms of the ion path lengths. Notably, the proximal rise of the annihilation distribution, even though not as steep as its distal fall-off, correlates with the location of the frontal edge of the SOBP. This all would ideally allow to spot the deviations in the dose distributions, associated with the range uncertainties, and correct them online.

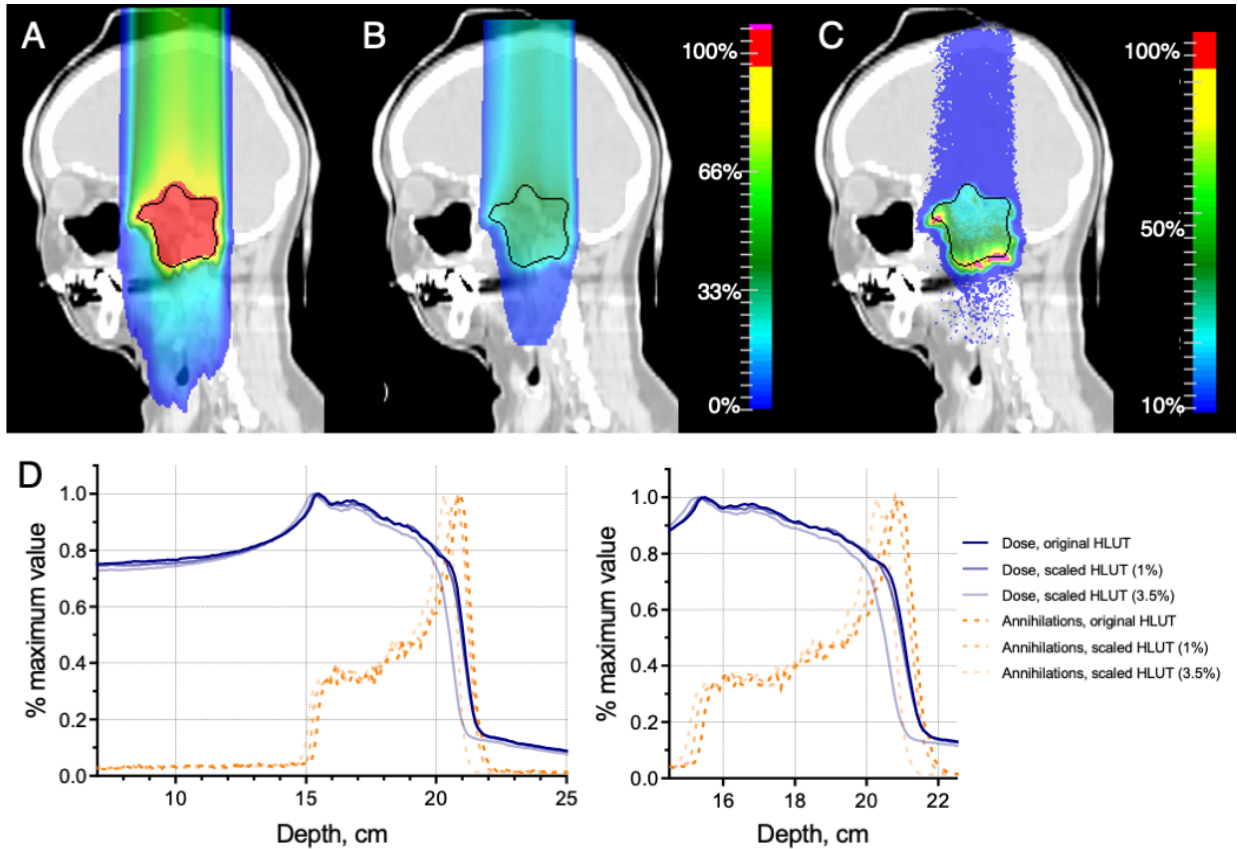

*Figure 1. RBE-weighted dose map (A) calculated with TRiP98, absorbed dose (B) and annihilation (C) maps calculated with FLUKA for a single-field $^{11}$C treatment plan optimized in a patient CT. Color bars reflect the respective distributions with red corresponding to the maximum observed values. (D) Normalized beam-eye-view profiles extracted from the dose and annihilation maps for the original treatment plan and those optimized using the scaled HLUT tables (for 3.5% or 1%) to mimic the range uncertainties. Depth is given in the CT coordinates. Blue solid lines: dose profiles, orange dashed lines – annihilation profiles. Left plot is a zoom-in of the distributions in the target region.*

### B. Margin reduction with $^{11}$C beams– impact on treatment plans

Following the outcomes of the preliminary study described in the previous section, we performed the robust biological (RBE-weighted) optimization of ACC and liver plans and the analysis of resulting dose distributions. Assuming the possibility of precise range verification of $^{11}$C beams with PET, the range uncertainty in the $^{11}$C plans has been set to 0%, while for the $^{12}$C plans the optimization and dose recalculation was done assuming the value of 3.5%. The setup uncertainty of ±3 mm has been included for both ions. This results in implicit range margins resulting from position shifts for both ions.

An example of the field configuration and organ contours for the ACC plan optimization is shown in Figure 2a. This information was preserved from the original plans used at GSI for patient treatment, where the nerves sometimes were bordering or overlapping with the CTVs. In these cases, since the priority was given to the target coverage ($D_{95} > 95\%$), sometimes the nerve(s) sparing could have been compromised.

Figure 2b summarizes patient by patient the treatment planning outcomes for both nerves patient by patient, i.e. the percentage of treatment scenarios (out of a total of 21) when the treatment planning constraint ($D_{0.03cm3} > 2.46$ Gy(RBE)) was violated to ensure the sufficient target coverage. The worst outcome would be 0% for both nerves, meaning that it is impossible to meet the planning constraint without compromising CTV coverage. Instead, the best outcome (100% for both nerves) would mean that sufficient target coverage will be achieved in all the treatment scenarios without exceeding the recommended doses for any nerve. The use of $^{11}$C plans with reduced range margins led to a significant improvement in the treatment plan quality. For only 2 patients out of 15 (in contrast to 8 out of 15 in the case of $^{12}$C), the maximal allowed dose was exceeded for both nerves in more than half of the analyzed scenarios.

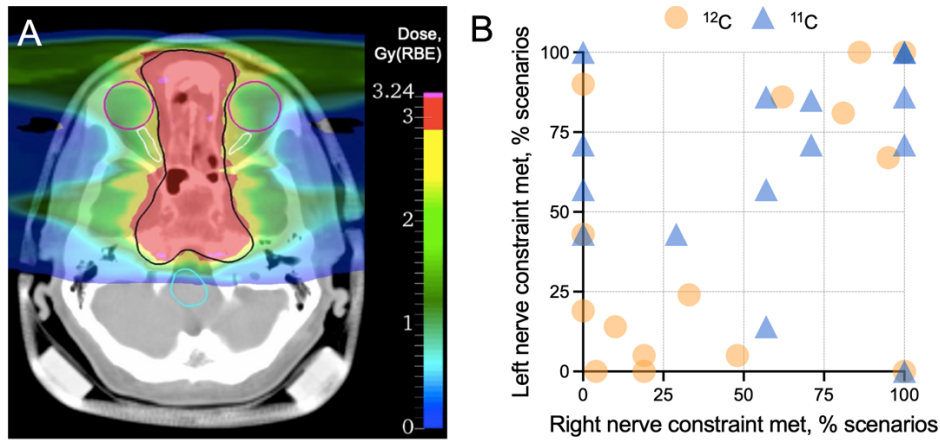

*Figure 2. Treatment planning outcome for the adenoid cystic carcinoma patients. (A) Example of the double-field $^{11}$C dose distribution and planning contours. Black – tumor CTV; magenta – eyes, white – optic nerves, turquise – brainstem. Color bar represents the dose distribution (red color corresponds to the prescribed CTV dose of 3 Gy(RBE)). (B) Patient-by-patient treatment planning outcome in terms of exceeding the dosimetric constraint for the optic nerves (2.46 Gy(RBE) to 0.03 cm³ of the organ). Each point corresponds to the percentage of treatment uncertainty scenarios when the constraint is met. Orange circles represent $^{12}$C plans (3 mm setup margin and 3.5% range margin), blue triangles – $^{11}$C plans (3 mm setup only).*

Figure 3a shows the example of the liver patient plan optimized for a uniform CTV (black contour) dose of 10 Gy(RBE). The liver is depicted with a brown contour. As in the ACC plans, the primary goal was achieving the prescribed target dose. The dose constraints set for the remaining healthy liver (as suggested by [52]), $D_{700cm^3}$ < 4.2 Gy(RBE) or 3.5 Gy(RBE) per fraction for 4- and 6-fraction schemes, respectively, were met for all the analyzed treatment plans.

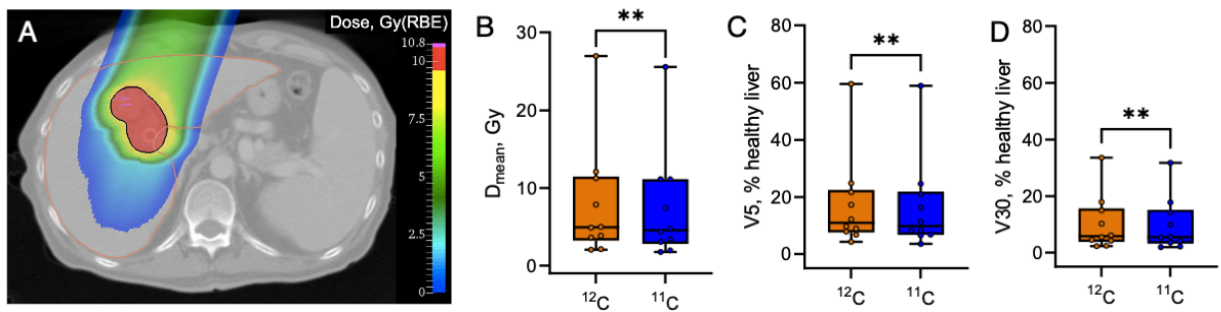

*Figure 3. Treatment outcomes planning for the liver tumor patients. (A) Example of the single-field dose distribution and planning contours. Black – tumor CTV, brown – liver. Color bar represents the dose distribution (red color corresponds to the prescribed CTV dose of 10 Gy(RBE)). (B) Total mean dose to the healthy liver (liver – GTV). (C) Relative volumes of healthy liver, receiving the total dose of at least 5 Gy (V5). (D) Relative volumes of healthy liver, receiving the total dose of at least 30 Gy (V30). Points represent the single patient median values among 21 uncertainty scenario (10 patients in total). Orange color: $^{12}$C plans, blue color: $^{11}$C plans. Stars represent the result of the Wilcoxon Signed-Rank test (P < 0.01) for both plots.*

Box plots in Figure 3b-d show the distribution of these parameters' median values from all the patients (i.e., 10 values per group). The outlier points in these plots correspond to the parameters derived for patient 6 with the largest CTV (223 cm$^3$), and the smallest healthy liver volumes (927 cm$^3$). Results of the Wilcoxon Signed-Rank reveal a significant difference (P < 0.01) for the $D_{mean}$, V5, and V30 values between $^{11}$C and $^{12}$C groups suggesting that also for the liver the use of $^{11}$C would be beneficial in terms of dose reduction.

### C. Influence of reduced margins on the estimated normal tissue toxicity

1. Serial organs: optic nerves

Box plots in Figure 4 depict the results of the NTCP calculations with the LKB model for the ACC patients with the endpoint of optic neuropathy for the right and left optic nerves. Each point represents the median NCTP value for a single patient over the 21 plan uncertainty scenarios. The median NTCP values for the left nerve for all the analyzed patients and scenarios are 2.5% vs. 1.1% for $^{11}$C and $^{12}$C plans, respectively; for the right nerve, these are 3.5% vs. 7.6%. The highest worst-case scenario NTCP values observed for $^{12}$C plans were 45.6% and 53.2% for the right and left nerves, respectively; in $^{11}$C, these values for the same patients dropped to 2.9% and 20.4%, respectively (not depicted here). The worst outcome in terms of NTCP was predicted for the patients where either the nerve was overlapping with the target or 'wrapped' by the tumor. Results of the Wilcoxon Signed-Rank reveal a significant difference (P < 0.0001) between the predicted NTCP values for $^{11}$C and $^{12}$C plans, suggesting that $^{11}$C plans with reduced range margins might be notably safer options compared to conventional $^{12}$C plans.

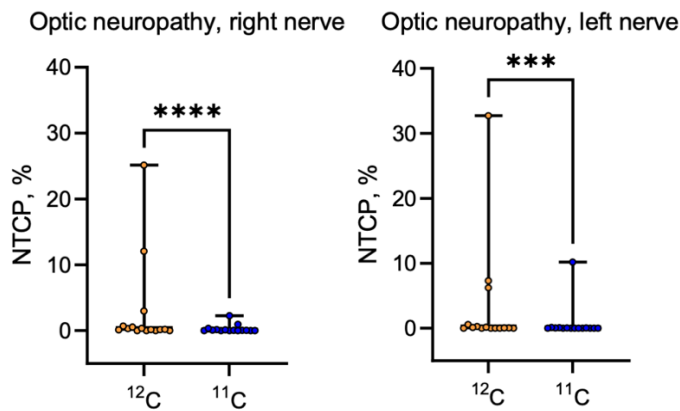

*Figure 4. Normal tissue complication probability (endpoint optic neuropathy) estimated for the right and left optic nerves for each patient using the Lyman-Kutcher-Burman model with parameters from [47] for $^{12}$C and $^{11}$C treatment plans. Single points correspond to the patient specific median NTCP values. Stars represent the result of the Wilcoxon Signed-Rank test (P < 0.0001) for both plots.*

We have additionally calculated the relative risks (RR) as RR = NTCP$_{11C}$/NTCP$_{12C}$ to quantify the relative NTCP difference between different ion plans for nominal and worst-case scenarios patient-by-patient. Figure 5 shows the RR values for both right (dashed bar) and left (solid color bar) nerves. RR < 1 would mean a safer outcome for $^{11}$C plans as compared to $^{12}$C plans. Despite the RR being >1 for three patients in nominal scenarios, in the worst-case scenarios the use of $^{11}$C with decreased margins might reduce the relative risk down to less than 0.2 in most cases.

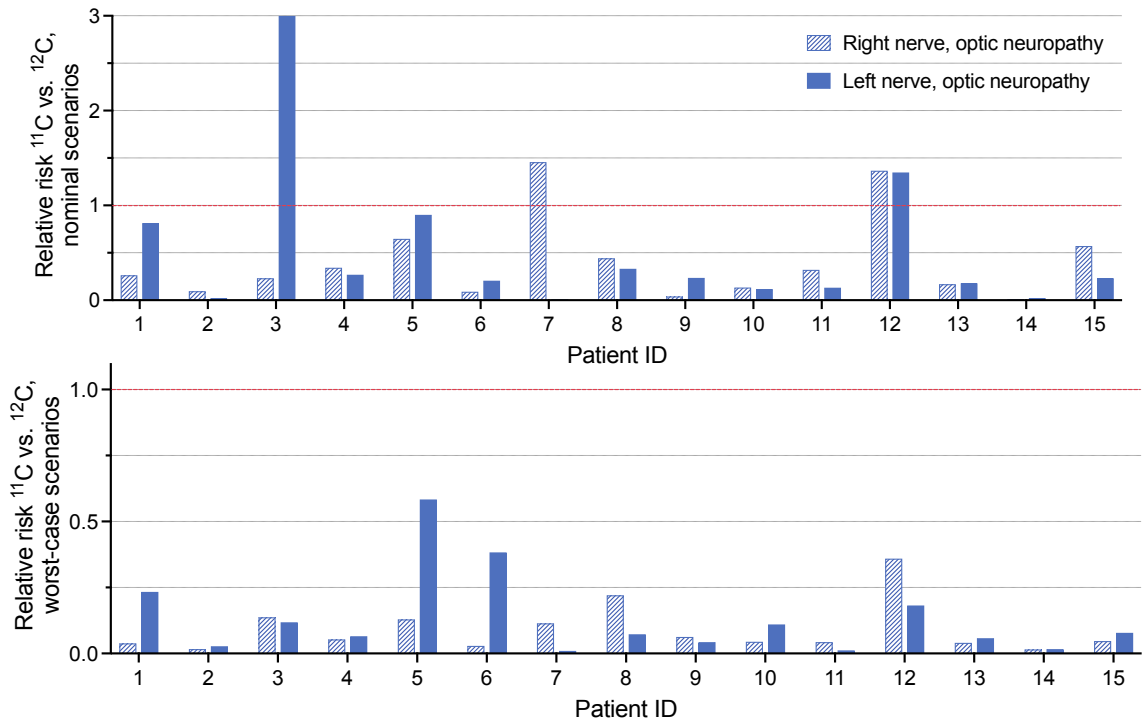

*Figure 5. Relative risks (RR) representing the ratio of $^{11}C$ vs. $^{12}C$ NTCP values for individual patients for the right (dashed bar) and left (solid color bars) optic nerves. RR < 1 means that the $^{11}C$ plan is beneficial compared to $^{12}C$ plan in terms of optic system toxicity. Top: nominal scenarios, bottom: worst-case scenarios.*

2. Parallel organs: liver

To estimate liver toxicity using the LKB model, we have used six models with different endpoints. The comparison of outcomes from $^{12}C$ vs. $^{11}C$ treatment plans is given in Figure 6. Apart from variations in the severity of the endpoints (from the liver failure (A) and the radiation-induced liver injury (B) to the potentially slightly less severe toxicities such as changes in albumin-bilirubin (ALBI) score, Child-Pugh (CP) score, or enzymatic changes (Figures C-F), the models also use different parameters as reported in Table 2. Of note, the NTCP predictions of the Burman model are quite varying from patient to patient and might reach quite high values; we attribute it to the low value of the volume effect parameter which increases the weight of the high-dose regions.

Wilcoxon Signed-Rank test reveals a significant difference ($P < 0.01$) between the NTCP values for $^{11}C$ and $^{12}C$ plans predicted by all six models. This suggests that $^{11}C$ plans with reduced range margins might be a better option also in terms of sparing the parallel critical structures.

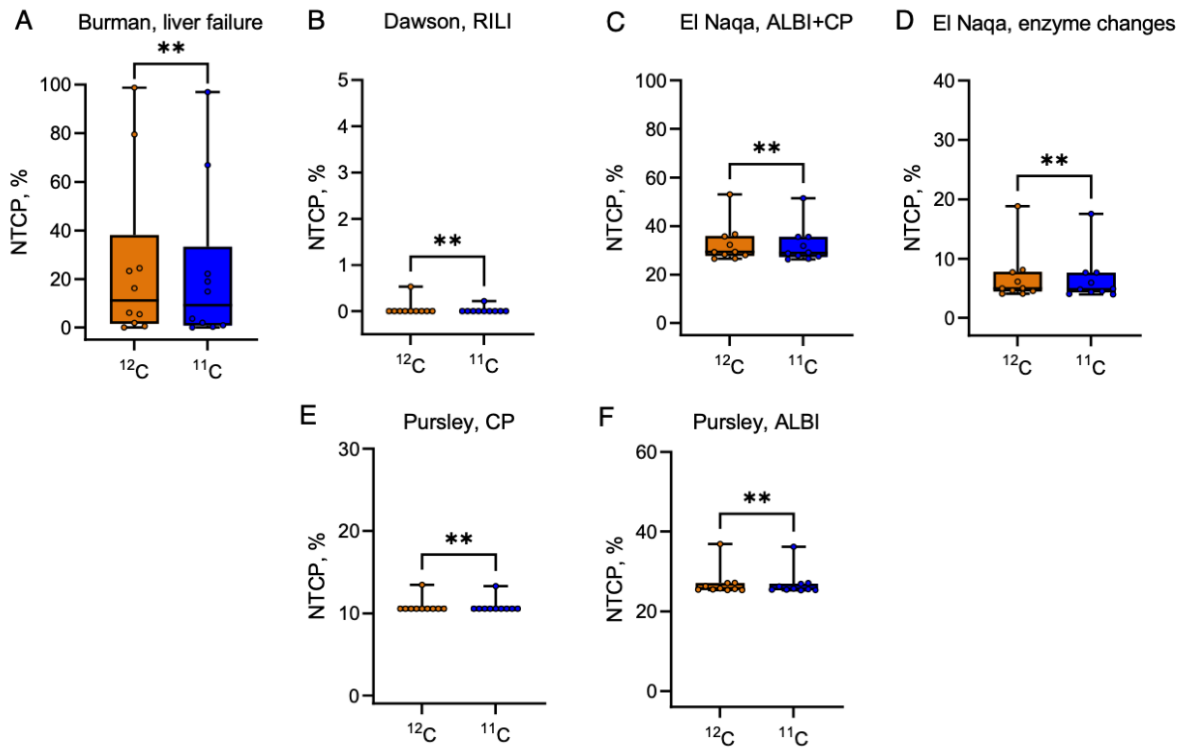

*Figure 6. Normal tissue complication probability estimated for each patient estimated using the LKB parameters from six different models. (A) Burman model for the endpoint of liver failure, (B) Dawson model for the endpoint of radiation-induced liver injury (RILI) grade > 3, (C) El Naqa model for changes in ALBI and CP scores, (D) El Naqa model for enzymatic changes, (E) Pursley model for changes in CP score, (F) Pursley model for changes in ALBI score. Single points correspond to the patient-specific median NTCP values. Orange and blue bars correspond to the predictions for $^{12}C$ and $^{11}C$ plans, respectively. Stars represent the result of the Wilcoxon Signed-Rank test (P < 0.01) for both plots.*

Similar to the ACC plan analysis, relative risks were calculated for an individual patient for every model and are presented in Figure 7 (top – nominal scenarios, bottom – worst-case scenarios). Depending on the model selected for comparison, $^{11}C$ plans would be either safer (Burman or Dawson models) than $^{12}C$ ones, or similar (El Naqua and Pursley models) in terms of relative risks. $^{11}C$ plans are most beneficial according to the Dawson model, where the RR decreases to 0.14 in the worst-case scenario for patient 7. However, one needs to keep in mind that for the Dawson model the absolute NTCP values are the lowest (less than 2%) compared to all the other models presented here.

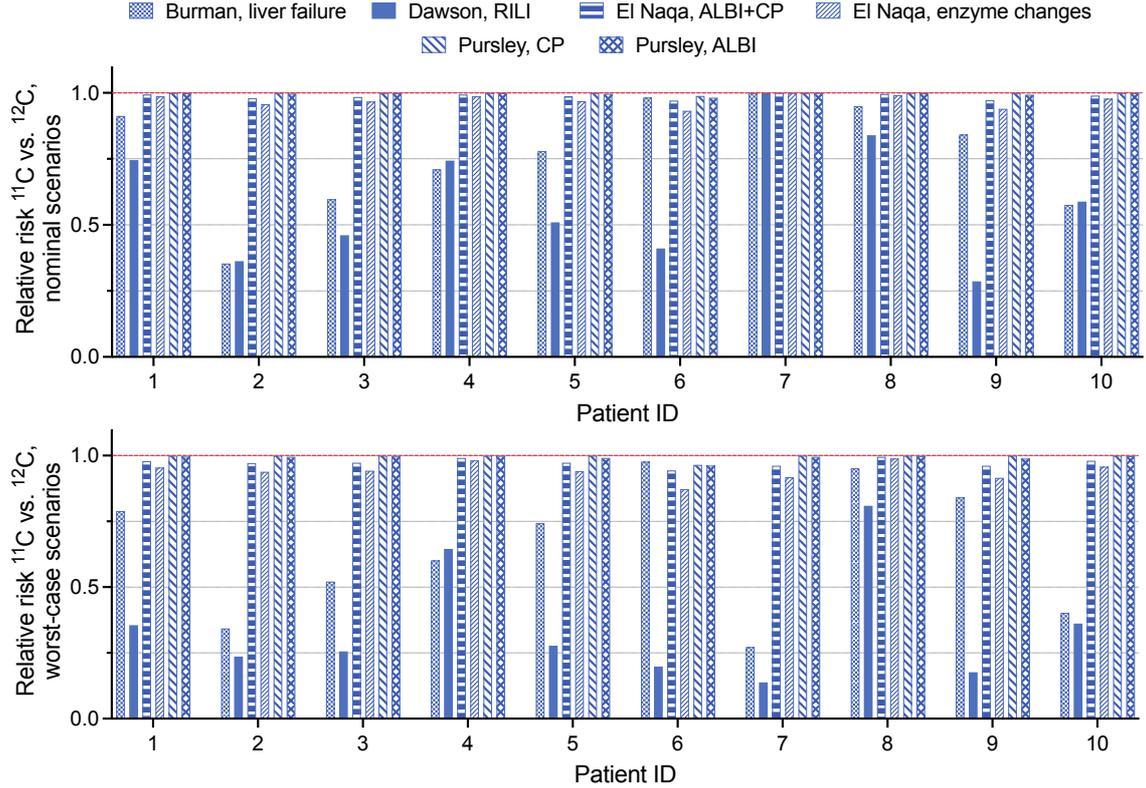

*Figure 7. Relative risk representing the ratio of $^{11}C$ vs. $^{12}C$ NTCP values for individual patients for each of the six models considered in the study. Value of RR < 1 means that the $^{11}C$ plan is beneficial compared to $^{12}C$ plan in terms of liver sparing. Top: nominal scenarios, bottom: worst-case scenarios.*

## IV. DISCUSSION

The results presented here show the potential benefits of the reduction of margins associated with range uncertainties from 3.5% to 0% for two different tumor types treated with carbon ions. This could be potentially achieved on a clinical level by using $^{11}C$ beams in combination with the improved PET range verification. In the previous overview of the BARB project [18], we have already presented a FLUKA comparison of the activity maps of $^{12}C$ and $^{11}C$ ion treatment plans in the simple geometry (sphere in a water tank). In this work with a new simulation study, we demonstrate that the annihilation map in the patient, irradiated with $^{11}C$ beams, can be used to determine the position of the SOBP distal fall-off and thus to recognize shifts in the dose distributions caused by the misinterpretation of the ion's water-equivalent pathlengths from CT data by 1%. This would mean that radioactive beams, in particular, $^{11}C$ studied here, would be an ideal tool for range monitoring during patient irradiation.

However, while the simulations presented here reflect the best possible scenario, where the whole annihilation signal could be used to generate PET images reflecting the shifts in the dose distributions caused by the range uncertainties, and a perfect scanner resolution, the practical feasibility of range margin complete elimination remains an open question. For example, the signal coming from $^{11}C$ beams with a half-life of about 20.33 min still might be affected by the washout that might slightly smear the distribution; from this point of view, $^{10}C$ beams might be more practical. Moreover, even if the look-up tables for the Hounsfield units would correctly reflect the ion's pathlengths, the precision of the Bragg peak position prediction in the patient would be limited by the CT resolution. Additionally, efficiency of the imaging detectors, the capability of the online image reconstruction software tools and the limited time used to accumulate the signal further limit the best possible achievable image quality. The results presented here should be thus interpreted as a 'best asymptotic' approximation of what could be achieved with RIB treatments.

It is important to point out that collecting the annihilation signal from the full plan delivered to the patient will be impractical if one aims at the online range verification since in this case the corrections can be applied only in the following fraction. The best approach to acquire enough signal in the shortest timespan possible to make any necessary plan modifications is yet to be identified. One option could be irradiating the whole target area but only at a small fraction of the target dose. Alternatively, a 'pilot shot' irradiating only a selected tumor slice with single energy would be faster since there is no time needed to switch between the different energies; however, there will be an increased risk of missing anatomical changes at higher depths.

Reduced range margins would make treatment planning procedures easier by increasing the chances of meeting the constraints set for the critical structures without compromising the desired target dose coverage. As the ACC plan analysis shows, with the standard margins, more than half of the plans were not acceptable in terms of exceeding the recommended maximal dose to the optic nerves in most of the treatment scenarios, while in the case of reduced margins the number of such 'problematic' plans is significantly reduced.

NTCP calculations for the optic nerves suggest the evident benefit of $^{11}$C plans with reduced margins for serial organs in target proximity. Apart from the reduction of the median value of NTCP estimated for each patient plan, the spread between the best and the worst-case scenarios was significantly reduced compared to $^{12}$C plans. Since the toxicity induction for the serial organ is rather the question of exceeding the certain dose in a point region, the range margin reduction, shrinking the high-dose region already by a millimeter, would play a crucial role in the organ sparing. Thus, the closer the serial OAR to the target structure, the more pronounced toxicity reduction should be expected. This agrees with the study [6] aimed at estimating the NTCP changes following the reduction of the range margins in proton therapy. To note, margin reduction in carbon therapy might have a stronger impact on the toxicity outcomes compared to protons, due to the 2-3-fold increase of RBE at the distal edge of the SOBP.

Concerning liver tumors, it is difficult to correlate the decrease of NTCP with some specifics of patient geometry and/or field configuration. While all the NTCP models applied here indeed predict the highest toxicities for the patients with bigger tumors and higher ratios of tumor volume to a healthy liver, to our surprise, the most decreased relative risks were estimated for the patients with smaller tumors.

The less pronounced effect of margin reduction for the liver plans compared to the ACC plans might be at least partially explained by the healthy liver volume definition. Treatment plans were optimized for a uniform CTV dose, whereas the healthy liver was defined as the total liver minus GTV, following standard recommendations [52]. That would mean that for both $^{11}$C and $^{12}$C plans a small fraction of the healthy liver (CTV – GTV) would always receive the target dose and thus the margin reduction would not have any pronounced impact on the maximal dose value received by the organ.

It is also important to note here, that when comparing liver plans in terms of the NTCP, attention needs to be paid to the differences in the models used. For example, in certain uncertainty scenarios for some patients, the Burman model for liver failure might predict the NTCP values higher than 60%, even though the treatment planning constraints were met. This might be attributed to the low value of the *n* parameter used in the respective LKB model which de-emphasizes the volumetric effects in the parallel organ and treats it more like a serial one. On the other hand, the Dawson model for the RILI predicts the NTCP values which are practically zero for all the patients, which might make the benefit of the five-time NTCP decrease with $^{11}$C beams questionable. Moreover, the models used here consider different endpoints, among which some of them might be considered more severe than others (e.g., liver failure vs. enzymatic changes). That is why we think that including all these models in the analysis would give a more complete picture. All in all, for the liver plans and NTCP models studied here, $^{11}$C plans reduce liver toxicity following the statistical analysis with the Wilcoxson test.

To further investigate which tumors will benefit the most from the margin reduction with RIBs, especially those close to the parallel organs, broader patient sets with different tumor sizes and their relative locations need to be analyzed. From the same point of view, considering other parallel organs, e.g., the kidney, would be interesting.

Reduced range uncertainties guaranteed with $^{11}$C beams, apart from reducing the doses to the adjacent structures by default, might further improve the flexibility and quality of treatment by enabling irradiation from different angles. A recent study [53] has shown further brainstem NTCP decrease from using novel proton beam arrangements, enabled by the reduced range margins. As the authors have shown, new beam arrangements might also enable the possibility of redistributing the LET distributions in the target and increasing its mean value. Concerning the $^{11}$C beams, the redistribution of the stronger LET gradients might be of further benefit for both sparing the OARs and tackling more radioresistant, e.g., hypoxic target regions [54–56].

In the present study, we did not account for target motion that would exacerbate the range uncertainty issue. For ACC tumors, this is a fair assumption. Instead, liver patients studied here were originally receiving SBRT with abdominal compression – for the heavy ion fields with sharp dose fall-offs, this mitigation technique would not be sufficient. Future studies should be thus extended to the 4D robust optimization to include additional motion-induced range changes.

## V. CONCLUSION

Radioactive heavy ion beams, such as $^{11}$C, can be ideal candidates to tackle the problem of range uncertainty in particle radiotherapy thanks to their sharp activity distribution that can be monitored using PET. In contrast to the PET signal produced by $^{12}$C ions, used nowadays in clinical practice, the signal from $^{11}$C would need a shorter acquisition time, and have a higher peak-to-noise ratio and a well-defined peak that can be better correlated with the dose distribution. All of this would allow *in vivo* beam monitoring in the patient. Improved precision of range monitoring would lead to smaller or eliminated range margins set for the target, which, in turn, would decrease the volume of normal tissue receiving unnecessary high doses of radiation. We have shown here that it is expected to have a positive impact on both the treatment planning outcomes (i.e., meeting treatment planning constraints without compromising the target dose coverage) and the predicted post-treatment toxicities (reduction of NTCP for various endpoints) both serial and parallel organs at risk in target proximity.

## ACKNOWLEDGMENTS

We would like to thank Dr. Moritz Work for the introduction into robust optimization in TRiP98 and Dr. Oliver Jäkel for help in interpreting the GSI patient data. The work is supported by European Research Council (ERC) Advanced Grant 883425 (BARB) to Marco Durante.

## REFERENCES

[1]  M. Durante, R. Orecchia, and J. S. Loeffler, *Charged-Particle Therapy in Cancer: Clinical Uses and Future Perspectives*, Nat Rev Clin Oncol **14**, 483 (2017).

[2]  L. Schaub, S. ben Harrabi, and J. Debus, *Particle Therapy in the Future of Precision Therapy*, Br J Radiol **93**, 20200183 (2020).

[3]  C. Grau, M. Durante, D. Georg, J. A. Langendijk, and D. C. Weber, *Particle Therapy in Europe*, Mol Oncol **14**, 1492 (2020).

[4]  M. Durante, J. Debus, and J. S. Loeffler, *Physics and Biomedical Challenges of Cancer Therapy with Accelerated Heavy Ions*, Nature Reviews Physics **3**, 777 (2021).

[5]  W. Tinganelli and M. Durante, *Carbon Ion Radiobiology*, Cancers (Basel) **12**, 3022 (2020).

[6]  S. Tattenberg, T. M. Madden, B. L. Gorissen, T. Bortfeld, K. Parodi, and J. Verburg, *Proton Range Uncertainty Reduction Benefits for Skull Base Tumors in Terms of Normal Tissue Complication Probability (NTCP) and Healthy Tissue Doses*, Med Phys **48**, 5356 (2021).

[7]  A. C. Kraan, *Range Verification Methods in Particle Therapy: Underlying Physics and Monte Carlo Modeling*, Front Oncol **5**, (2015).

[8]  F. Sommerer, F. Cerutti, K. Parodi, A. Ferrari, W. Enghardt, and H. Aiginger, *In-Beam PET Monitoring of Mono-Energetic $^{16}O$ and $^{12}C$ Beams: Experiments and FLUKA Simulations for Homogeneous Targets*, Phys Med Biol **54**, 3979 (2009).

[9]  H. J. T. Buitenhuis, F. Diblen, K. W. Brzezinski, S. Brandenburg, and P. Dendooven, *Beam-on Imaging of Short-Lived Positron Emitters during Proton Therapy*, Phys Med Biol **62**, 4654 (2017).

[10] A. C. Kraan et al., *Proton Range Monitoring with In-Beam PET: Monte Carlo Activity Predictions and Comparison with Cyclotron Data*, Physica Medica **30**, 559 (2014).

[11] J. Handrack, T. Tessonnier, W. Chen, J. Liebl, J. Debus, J. Bauer, and K. Parodi, *Sensitivity of Post Treatment Positron Emission Tomography/Computed Tomography to Detect Inter-Fractional Range Variations in Scanned Ion Beam Therapy*, Acta Oncol (Madr) **56**, 1451 (2017).

[12] S. P. Nischwitz et al., *Clinical Implementation and Range Evaluation of in Vivo PET Dosimetry for Particle Irradiation in Patients with Primary Glioma*, Radiotherapy and Oncology **115**, 179 (2015).

[13] M. Durante and K. Parodi, *Radioactive Beams in Particle Therapy: Past, Present, and Future*, Front. Phys. **8**, 326 (2020).

[14] R. S. Augusto et al., *An Overview of Recent Developments in FLUKA PET Tools*, Physica Medica **54**, 189 (2018).

[15] C. Toramatsu et al., *Washout Effect in Rabbit Brain: In-Beam PET Measurements Using $^{10}C$, $^{11}C$ and $^{15}O$ Ion Beams*, Biomed Phys Eng Express **4**, 035001 (2018).

[16] A. Mohammadi et al., *Influence of Momentum Acceptance on Range Monitoring of $^{11}C$ and $^{15}O$ Ion Beams Using in-Beam PET*, Phys Med Biol **65**, 125006 (2020).

[17] M. Durante et al., *All the Fun of the FAIR: Fundamental Physics at the Facility for Antiproton and Ion Research*, Phys Scr **94**, 033001 (2019).

[18] D. Boscolo et al., *Radioactive Beams for Image-Guided Particle Therapy: The BARB Experiment at GSI*, Front. Oncol. **11**, 737 (2021).

[19] D. Boscolo et al., *Depth Dose Measurements in Water for 11C and 10C Beams with Therapy Relevant Energies*, Nucl Instrum Methods Phys Res A 167464 (2022).

[20] G. Palma, S. Monti, M. Conson, R. Pacelli, and L. Cella, *Normal Tissue Complication Probability (NTCP) Models for Modern Radiation Therapy*, Semin Oncol **46**, 210 (2019).

[21] A. D. Jensen, A. v. Nikoghosyan, M. Poulakis, A. Höss, T. Haberer, O. Jäkel, M. W. Münter, D. Schulz-Ertner, P. E. Huber, and J. Debus, *Combined Intensity-Modulated Radiotherapy plus Raster-Scanned Carbon Ion Boost for Advanced Adenoid Cystic Carcinoma of the Head and Neck Results in Superior Locoregional Control and Overall Survival*, Cancer **121**, 3001 (2015).

[22] A. D. Jensen et al., *High-LET Radiotherapy for Adenoid Cystic Carcinoma of the Head and Neck: 15 Years' Experience with Raster-Scanned Carbon Ion Therapy*, Radiotherapy and Oncology **118**, 272 (2016).

[23] A. Gemmel, B. Hasch, M. Ellerbrock, W. Kraft-Weyrather, and M. Krämer, *Biological Dose Optimization with Multiple Ion Fields*, Phys Med **53**, 6691 (2008).

[24] M. Wolf, K. Anderle, M. Durante, and C. Graeff, *Robust Treatment Planning with 4D Intensity Modulated Carbon Ion Therapy for Multiple Targets in Stage IV Non-Small Cell Lung Cancer*, Phys Med Biol **65**, 215012 (2020).

[25] W. Liu, S. J. Frank, X. Li, Y. Li, P. C. Park, L. Dong, X. Ronald Zhu, and R. Mohan, *Effectiveness of Robust Optimization in Intensity-Modulated Proton Therapy Planning for Head and Neck Cancers*, Med Phys **40**, 051711 (2013).

[26] J. Schuemann, S. Dowdell, C. Grassberger, C. H. Min, and H. Paganetti, *Site-Specific Range Uncertainties Caused by Dose Calculation Algorithms for Proton Therapy*, Phys Med Biol **59**, 4007 (2014).

[27] M. Lambrecht et al., *Radiation Dose Constraints for Organs at Risk in Neuro-Oncology; the European Particle Therapy Network Consensus*, Radiotherapy and Oncology **128**, 26 (2018).

[28] M. C. Joiner and A. J. van der Kogel, editors, *Basic Clinical Radiobiology* (CRC press, 2018).

[29] S. H. Benedict et al., *Stereotactic Body Radiation Therapy: The Report of AAPM Task Group 101*, Med Phys **37**, 4078 (2010).

[30] J. Wulf, M. Guckenberger, U. Haedinger, U. Oppitz, G. Mueller, K. Baier, and M. Flentje, *Stereotactic Radiotherapy of Primary Liver Cancer and Hepatic Metastases*, Acta Oncol (Madr) **45**, 838 (2006).

[31] K. Shibuya, T. Ohno, H. Katoh, M. Okamoto, S. Shiba, Y. Koyama, S. Kakizaki, K. Shirabe, and T. Nakano, *A Feasibility Study of High-Dose Hypofractionated Carbon Ion Radiation Therapy Using Four Fractions for Localized Hepatocellular Carcinoma Measuring 3 cm or Larger*, Radiotherapy and Oncology **132**, 230 (2019).

[32] L. A. Dawson, D. Normolle, J. M. Balter, C. J. McGinn, T. S. Lawrence, and R. K. ten Haken, *Analysis of Radiation-Induced Liver Disease Using the Lyman NTCP Model*, International Journal of Radiation Oncology*Biology*Physics **53**, 810 (2002).

[33] R. Michel, I. Françoise, P. Laure, M. Anouchka, P. Guillaume, and K. Sylvain, *Dose to Organ at Risk and Dose Prescription in Liver SBRT*, Reports of Practical Oncology & Radiotherapy **22**, 96 (2017).

[34] M. Krämer, O. Jäkel, T. Haberer, D. Schardt, and U. Weber, *Treatment Planning for Heavy-Ion Radiotherapy: Physical Beam Model and Dose Optimization*, Phys Med Biol **3299**, (2000).

[35] G. Battistoni et al., *The FLUKA Code: An Accurate Simulation Tool for Particle Therapy*, Front. Oncol. **6**, 116 (2016).

[36] K. Parodi, A. Mairani, S. Brons, B. G. Hasch, F. Sommerer, J. Naumann, O. Jäkel, T. Haberer, and J. Debus, *Monte Carlo Simulations to Support Start-up and Treatment Planning of Scanned Proton and Carbon Ion Therapy at a Synchrotron-Based Facility*, Phys Med Biol **57**, 3759 (2012).

[37] K. Parodi, A. Mairani, S. Brons, B. G. Hasch, F. Sommerer, J. Naumann, O. Jäkel, T. Haberer, and J. Debus, *Monte Carlo Simulations to Support Start-up and Treatment Planning of Scanned Proton and Carbon Ion Therapy at a Synchrotron-Based Facility*, Phys Med Biol **57**, 3759 (2012).

[38] A. Fedorov et al., *3D Slicer as an Image Computing Platform for the Quantitative Imaging Network*, Magn Reson Imaging **30**, 1323 (2012).

[39] G. Palma, S. Monti, A. Buonanno, R. Pacelli, and L. Cella, *PACE: A Probabilistic Atlas for Normal Tissue Complication Estimation in Radiation Oncology*, Front Oncol **9**, (2019).

[40] G. Palma, S. Monti, and L. Cella, *Voxel-Based Analysis in Radiation Oncology: A Methodological Cookbook*, Physica Medica **69**, 192 (2020).

[41] J. D. Chapman and A. E. Nahum, *Radiotherapy Treatment Planning* (CRC Press, 2016).

[42] M. Astrahan, *Some Implications of Linear-Quadratic-Linear Radiation Dose-Response with Regard to Hypofractionation*, Med Phys **35**, 4161 (2008).

[43] I. el Naqa, A. Johansson, D. Owen, K. Cuneo, Y. Cao, M. Matuszak, L. Bazzi, T. S. Lawrence, and R. K. ten Haken, *Modeling of Normal Tissue Complications Using Imaging and Biomarkers After Radiation Therapy for Hepatocellular Carcinoma*, International Journal of Radiation Oncology*Biology*Physics **100**, 335 (2018).

[44] J. Pursley et al., *Dosimetric Analysis and Normal-Tissue Complication Probability Modeling of Child-Pugh Score and Albumin-Bilirubin Grade Increase After Hepatic Irradiation*, International Journal of Radiation Oncology*Biology*Physics **107**, 986 (2020).

[45] C. Burman, G. J. Kutcher, B. Emami, and M. Goitein, *Fitting of Normal Tissue Tolerance Data to an Analytic Function*, International Journal of Radiation Oncology*Biology*Physics **21**, 123 (1991).

[46] A. Niemierko, *A Generalized Concept of Equivalent Uniform Dose (EUD)*, Med Phys **26**, (1999).

[47] C. Mayo, E. Yorke, and T. E. Merchant, *Radiation Associated Brainstem Injury*, International Journal of Radiation Oncology*Biology*Physics **76**, S36 (2010).

[48] C. Burman, G. J. Kutcher, B. Emami, and M. Goitein, *Fitting of Normal Tissue Tolerance Data to an Analytic Function*, International Journal of Radiation Oncology*Biology*Physics **21**, 123 (1991).

[49] L. A. Dawson, D. Normolle, J. M. Balter, C. J. McGinn, T. S. Lawrence, and R. K. ten Haken, *Analysis of Radiation-Induced Liver Disease Using the Lyman NTCP Model*, International Journal of Radiation Oncology*Biology*Physics **53**, 810 (2002).

[50] M. Velec et al., *Predictors of Liver Toxicity Following Stereotactic Body Radiation Therapy for Hepatocellular Carcinoma*, International Journal of Radiation Oncology*Biology*Physics **97**, 939 (2017).

[51] P.-C. Shen, W.-Y. Huang, Y.-H. Dai, C.-H. Lo, J.-F. Yang, Y.-F. Su, Y.-F. Wang, C.-F. Lu, and C.-S. Lin, *Radiomics-Based Predictive Model of Radiation-Induced Liver Disease in Hepatocellular Carcinoma Patients Receiving Stereo-Tactic Body Radiotherapy*, Biomedicines **10**, 597 (2022).

[52] M. Miften et al., *Radiation Dose-Volume Effects for Liver SBRT*, International Journal of Radiation Oncology*Biology*Physics **110**, 196 (2021).

[53] S. Tattenberg, T. M. Madden, T. Bortfeld, K. Parodi, and J. Verburg, *Range Uncertainty Reductions in Proton Therapy May Lead to the Feasibility of Novel Beam Arrangements Which Improve Organ-at-risk Sparing*, Med Phys **49**, 4693 (2022).

[54] N. Bassler, J. Toftegaard, A. Lühr, B. S. Sørensen, E. Scifoni, M. Krämer, O. Jäkel, L. S. Mortensen, J. Overgaard, and J. B. Petersen, *LET-Painting Increases Tumour Control Probability in Hypoxic Tumours*, Acta Oncol (Madr) **53**, 25 (2014).

[55] E. Malinen and Å. Søvik, *Dose or 'LET' Painting – What Is Optimal in Particle Therapy of Hypoxic Tumors?*, Acta Oncol (Madr) **54**, 1614 (2015).

[56] W. Tinganelli, M. Durante, R. Hirayama, M. Krämer, A. Maier, W. Kraft-Weyrather, Y. Furusawa, T. Friedrich, and E. Scifoni, *Kill-Painting of Hypoxic Tumours in Charged Particle Therapy.*, Sci Rep **5**, 1 (2015).